# Towards an IT Security Risk Assessment Framework for Railway Automation

Jens Braband

Siemens AG, Braunschweig, Germany
jens.braband@siemens.com

**Abstract.** Some recent incidents have shown that possibly the vulnerability of IT systems in railway automation has been underestimated. Fortunately, so far, almost only denial-of-service attacks were successful, but due to several trends, such as the use of commercial IT and communication systems or privatization, the threat potential could increase in the near future. However, up to now, no harmonized IT security risk assessment framework for railway automation exists. This paper defines an IT security risk assessment framework which aims to separate IT security and safety requirements as well as certification processes as far as possible. It builds on the well-known safety and approval processes from IEC 62425 and integrates IT security requirements based on the ISA99/IEC62443 standard series. While the detailed results are related to railway automation the general concepts are also applicable to other safety-critical application areas.

**Keywords.** Railway, IT Security, Safety, Risk Assessment, IT Security Requirements.

## 1    Introduction

Over the last years, reports on IT security incidents related to railways have increased as well as public awareness. For example, it was reported that, on December 1, 2011, "hackers, possibly from abroad, executed an attack on a Northwest rail company's computers that disrupted railway signals for two days" [1]. Although the details of the attack and also its consequences remain unclear, this episode clearly shows the threats to which railways are exposed when they rely on modern commercial-off-the-shelf (COTS) communication and computing technology. However, in most cases, the attacks are denial-of-service attacks leading to service interruptions, but so far not to safety-critical incidents. But also other services, such as satellite positioning systems, have been shown to be susceptible to IT security attacks, leading to a recommendation that GNSS services should not be used as standalone positioning services for safety-related applications [2].

What distinguishes railway systems from many other systems is their inherently distributed and networked nature with tens of thousands of kilometer track length for large operators. Thus, it is not economical to provide complete protection against physical access to this infrastructure and, as a consequence, railways are very vulnerable to physical denial-of-service attacks leading to service interruptions.

Another feature of railways distinguishing them from most other systems is the long lifespan of their systems and components. Current contracts usually demand support for over 25 years and history has shown that many systems, e.g. mechanical or relay interlockings, last much longer. IT security analyses have to take into account such a long lifespan. Nevertheless, it should also be noted that at least some of the technical problems are not railway-specific, but are shared by other sectors such as Air Traffic Management [3].

Concerning IT security another difference to many other application sectors is that railway automation is a highly safety-critical field, which has a rather strict approval regime similar to civil aviation. It seems that so far many IT security considerations have been made without this background. While in railway automation harmonized safety standards were elaborated almost two decades ago, up to now no harmonized IT security requirements for railway automation exist.

This paper starts with a discussion of the normative and legal background. A short overview of the basic concepts of ISA99/IEC62443 [4] is given. Then several approaches towards IT security risk assessment are discussed with particular focus on their applicability to safety-critical systems. Then an IT security risk assessment framework is defined which aims to separate IT security and safety requirements as well as certification processes as far as possible. It is finally discussed how these concepts can be applied effectively to railway automation as well as other safety-critical domains.

## 2 Normative Background

In railway automation, there exists an established standard for safety-related communication, IEC 62280 [5]. The first version of the standard was elaborated in 2001. It has proven quite successful and is also used in other application areas, e.g. industry automation. This standard defines threats and countermeasures to ensure safe communication in railway systems. So, at an early stage, the standard established methods to build a safe channel (in security, called "tunnel" or "conduit") through an unsafe environment. However, the threats considered in IEC 62280 arise from technical sources or the environment rather than from humans. The methods described in the standard are partially able to protect the railway system also from intentional attacks, but not completely. Until now, additional organizational and technical measures have been implemented in railway systems, such as separated networks, etc., to achieve a sufficient level of protection.

The safety aspects of electronic hardware and systems are covered by IEC 62425 [6]. However, security issues are taken into account by IEC 62425 only as far as they affect safety issues, but, for example, denial-of-service attacks often do not fall into this category. Questions such as intrusion protection are only covered by one requirement in Table E.10 (unauthorized access). Nevertheless, IEC 62425 provides a structure for a safety case which explicitly includes a subsection on protection against unauthorized access (both physical and informational).

On the other hand, industrial standards on information security exist. Here, we can identify the following standards:

- ISO/IEC 15408 [7] provides evaluation criteria for IT security, the so-called Common Criteria [8,9,10]. This standard is solely centered on information systems and has, of course, no direct relation to safety systems.

- The ISA99/IEC62443 series is a set of standards currently elaborated by the Industrial Automation and Control System Security Committee of the International Society for Automation (ISA). This standard is not railway-specific and focuses on industrial control systems. It is dedicated to different hierarchical levels, starting from concepts and going down to components of control systems.

Railways are certainly critical national and international infrastructures, so recently national governments, e.g. the USA and Germany, as well as the EU have identified the problem. They have defined clear policies to support the implementation of industry-defined sector-specific IT security standards.

How can the gap between information security standards for general systems and railways be bridged? One bridge is provided by the European Commission Regulation No. 402/2013 on Common Safety Methods [11]. This Commission Regulation mentions three different methods to demonstrate that a railway system is sufficiently safe:

a)      by following existing rules and standards (application of codes of practice),

b)      by similarity analysis, i.e. showing that the given (railway) system is equivalent to an existing and used one,

c)      by explicit risk analysis, where risk is assessed explicitly and shown to be acceptable.

We assume that, from the process point of view, security can be treated just like safety, meaning that threats would be treated as particular hazards. Using the approach under a), Common Criteria  or ISA99/IEC62443 may be used in railway systems, but particular tailoring would have to be performed due to different safety requirements and application conditions. By this approach, a code of practice that is approved in other areas of technology and provides a sufficient level of security can be adapted to railways. This ensures a sufficient level of safety.

However, application of the general standards [4,7] requires tailoring them to the specific needs of a railway system. This is necessary to cover the specific threats associated with railway systems and possible accidents and to take into account specific other risk-reducing measures already present in railway systems, such as the use of specifically trained personnel.

This finally leads to a kind of "IT security for safety approach", where the IT security objectives and processes are referenced by  the technical safety report from IEC 62425, see Figure 1. Other security objectives can also be described in that structure, however the puzzle is not complete today and needs further railway-specific supporting standards and guidelines.

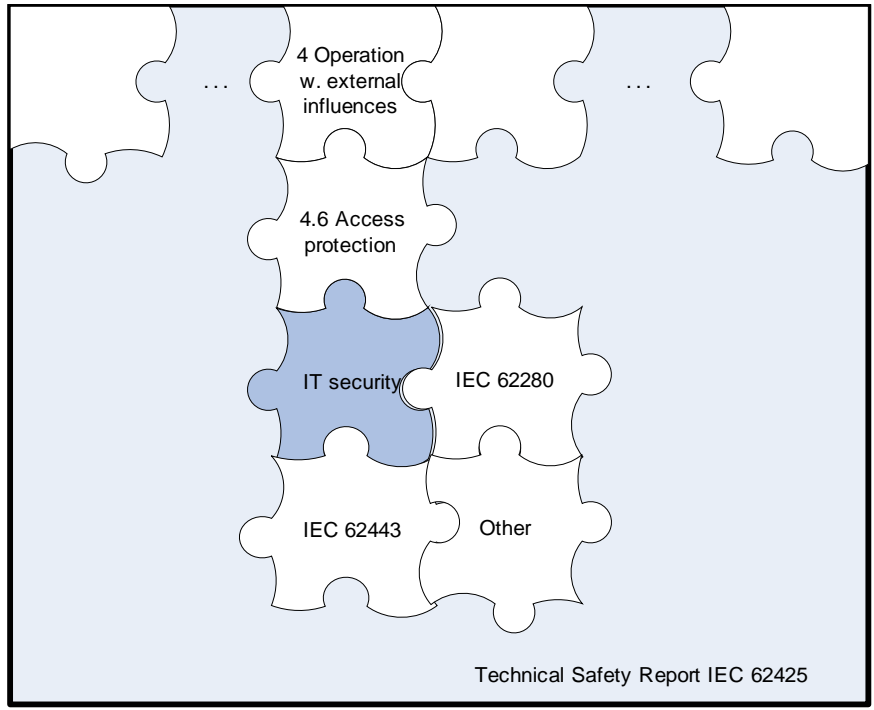

Figure 1: Embedding on IT security in the technical safety report from IEC 62425

# 3 Problems with Threat&Risk Analysis for Safety-related Systems

From the risk analysis point of view, many concepts from safety and IT security seem very similar; only the wording seems different. What's called a hazard in safety is called a threat in IT security, but the risk analysis processes really look alike. Thus it would be a logic conclusion to apply the same risk assessment techniques to IT security. This idea is even more supported by fact that many measures of IT security are adapted for safety and residual failure probabilities are computed, e.g. for transmission errors in communication, see e.g. IEC 62280. As a matter of fact this similarity is used in many security standards, ISO 27005 [12] being the most general, but instead of probability the term likelihood is introduced. A commonly used IT security risk matrix is shown in Figure 1.

| | Likelihood of incident scenario | Very Low (Very Unlikely) | Low (Unlikely) | Medium (Possible) | High (Likely) | Very High (Frequent) |
|---|---|---|---|---|---|---|
| | Very Low | 0 | 1 | 2 | 3 | 4 |
| | Low | 1 | 2 | 3 | 4 | 5 |
| Business Impact | Medium | 2 | 3 | 4 | 5 | 6 |
| | High | 3 | 4 | 5 | 6 | 7 |
| | Very High | 4 | 5 | 6 | 7 | 8 |

Figure 2: Risk matrix based on ISO 27005

In ISO 27005 "likelihood is used instead of the term 'probability' for risk estimation". It is admitted that "its ease of understanding" is an advantage, but "the dependence on subjective choice of scale" is a disadvantage.

NIST guidance [13] explains in more detail: "The likelihood of occurrence is a weighted risk factor based on an analysis of the probability that a given threat is capable of exploiting a given vulnerability (or set of vulnerabilities). The likelihood risk factor combines an estimate of the likelihood that the threat event will be initiated with an estimate of the likelihood of impact (i.e., the likelihood that the threat event results in adverse impacts). For adversarial threats, an assessment of likelihood of occurrence is typically based on: (i) adversary intent; (ii) adversary capability; and (iii) adversary targeting."

Aven [14] gives a broad discussion how all the different concepts can be combined in a risk analysis. However we have to distinguish here between general or business risks and safety risks. In economic risk assessments the goal is to find the best solution with a high probability under uncertainty. In safety risk assessments we have to take into account the fact that lives and health of people are at stake and that finally the risk assessments have to be accepted by a safety authority. In safety-critical systems this approach creates problems as it is hard to deal with subjective probabilities in the approval process.

Also scientifically the probabilistic approach does not really apply to IT security, e. g. we can't rely on statistical data or experience as we often do in safety as the threat landscape and risk assessment may change immediately if a new vulnerability becomes known. Also the attacks, at least not the targeted attacks, don't occur randomly. So as a matter of fact we have both systematic causes as the initiators of a security threat and we have vulnerabilities, flaws in the system or SW engineering, as contributing factors, which also have a systematic nature.

## 4 Overview of ISA99/IEC62443 Standards

Currently, 10 parts are planned in this standard series covering different aspects for industrial automation and control systems (IACS, the main stakeholders are addressed in brackets):

General (all):
- 1-1 Terminology, concepts and models
- 1-2 Master glossary of terms and abbreviations

Policies and procedures (railway operators):
- 2-1 Establishing an industrial automation and control system security program
  2-3 Patch management in the IACS environment
- 2-4 Security program equirements for IACS solution suppliers

System (system integrators):
- 3-1 Security technologies for IACS
- 3-2 Security levels for zones and conduits
- 3-3 System security requirements and security levels

Components (suppliers of IT security products)
- 4-1 Product development requirements
- 4-2 Technical security requirements for IACS products

The documents are at different stages of development, some being already international standards, while others are at the first drafting stage. This leads in particular to problems when the documents build on each other, e.g. Part 3-3 [15] with detailed security requirements is published, but it builds on Part 3-2 which defines the security levels and is restarted after a negative vote.

The fundamental concept of the standard is to define foundational requirements (FR) and security levels (SL) which are a "measure of confidence that the IACS is free from vulnerabilities and functions in the intended manner". There are seven groups of FR:

1. Identification and authentication control
2. Use control
3. System integrity
4. Data confidentiality
5. Restricted data flow
6. Timely response to events
7. Resource availability

Each FR group has up to 13 sub-requirement categories which are tailored according to the SL.

## 5    Security Levels

The default SL assignment for each zone and conduit is based on the attacker capability only:

SL1: casual or unintended
SL 2: simple means: low resources, generic skills and low motivation
SL 3: sophisticated means: moderate resources, IACS-specific skills and moderate motivation
SL 4: sophisticated means: extended resources, IACS-specific skills and high motivation

The default assignment can be changed based on the results of a threat and risk analysis. For each FR, a different SL may be assigned. There is a distinction between a target SL (as derived by threat and risk analysis), a design SL (capability of the solution architecture) and finally the achieved SL (as finally realized). If either the design SL or the achieved SL does not match the target SL, then additional measures have to be implemented (e.g. physical or organizational) as compensation.

Taking into account the fact that there may also be no IT security requirement (SL 0), an SL assignment results in a seven-dimensional vector with $5^7$=78.125 possible different assignments. Based on the SL assignment, a standardized set of IT security requirements can be found in Part 3-3, which is a great advantage of the approach.

Note that there is no simple match between SL and the Safety Integrity Levels (SIL) applied in safety standards. However, the definition of SL1 is very similar to requirements in the safety field as also safety-related systems have to address topics such as operator error, foreseeable misuse or effects of random failure. So we can conclude that a safety-related system (SIL>0) should also fulfill the technical requirements of IEC 62443-3-3 as the threats which SL1 addresses are also safety hazards.

A concise comparison shows that there are some differences in detail. IEC 62443-3-3 contains 41 requirements for SL1, of which more than half are directly covered by safety standards such as IEC 62425 or IEC 62280, and about a quarter are usually fulfilled in railway safety systems representing good practice. However, another quarter of the requirements are usually not directly addressed in IEC 62425 safety cases. The main reasons are that these requirements do not fall under the "IT security for safety" category but address availability requirements in order to prevent denial of service or traceability requirements.

The current proposal is to include all SL1 requirements from IEC 62443-3-3 in the system requirements specification of any safety-related signaling system. In this way, no additional SL1 IT security certification would be necessary and it would be a contribution to the defense-in-depth principle. Finally, these requirements should find their place in the IEC 62278 standards series.

For the sake of brevity, we are focusing on system aspects in this paper. The first step after system definition would be to divide the system into zones and conduits according to the following basic rules:

- The system definition must include all hardware and software objects.
- Each object is allocated to a zone or a conduit.
- Inside each zone, the same IT security requirements are applicable.
- There exists at least one conduit for communication with the environment.

The next step is the threat and risk analysis resulting in SL assignment to each zone and conduit. Here, railway applications might need procedures different from industry automation as factories and plants are usually physically well protected and are not moving.

As soon as the SL is assigned, standardized requirements from IEC 62443-3-3 can be derived. These requirements would be taken over to the railway automation domain without any technical changes. They would define the interface to use pre-certified IT security components for the railway automation domain.

Finally, correct implementation of the IT security countermeasures according to IEC 62443-3-3 must be evaluated similar to the validation of safety functions.

# 6 Approaches towards IT Security Risk Assessment

## 6.1 IEC 62443-3-2 proposal

Recently, a novel approach towards semi-quantitative IT security risk assessment has been proposed in the draft IEC 62443-3-2 [16]. But it did not pass the voting and so the work had to be restarted. However it is very interesting to study this approach and why it failed.

The new approach starts with a sample risk matrix as shown in figure 3. It looks like a common approach to determine the risk R for a particular threat from the parameters likelihood L and impact I by

$$R = L \cdot I \qquad (1)$$

| | | Likelihood | | | | |
|---|---|---|---|---|---|---|
| | | 1 Remote | 2 Unlikely | 3 Possible | 4 Likely | 5 Certain |
| **Impact** | 1 Trivial | 1 | 2 | 3 | 4 | 5 |
| | 2 Minor | 2 | 4 | 6 | 8 | 10 |
| | 3 Moderate | 3 | 6 | 9 | 12 | 15 |
| | 4 Major | 4 | 8 | 12 | 16 | 20 |
| | 5 Critical | 5 | 10 | 15 | 20 | 25 |

Figure 3: Sample risk matrix

Then, 4 is stipulated as a tolerable risk (without any justification) and any identified risk value R is then divided by 4, giving the so-called cyber security risk reduction factor (CRRF)

$$CRRF = \frac{R}{4} \qquad (2)$$

and finally the target SL is derived by

$$SL\text{-}T = \min\left\{4, \left\lceil CRRF\text{-}\frac{1}{4}\right\rceil\right\} \qquad (3)$$

Formula (3) simply states that a SL-T must not be larger than 4 and that it is more or less given by the integer part of the CRRF with a small correction of ¼. In order to understand it better, let us look at some interesting examples. For R=16, the CRRF is 4, which by (3) leads to SL-T=3. For R=17, it would lead to SL-T=4. Interestingly, both risks belong to the highest risk category in Table 1. Also, other border cases are interesting, e.g. risks labeled 6, 7 and 8 lead to SL-T=1, while 9 and 10 would result in SL-T=2. While all low-level risks should normally be acceptable, risks with 1, 2, 3, and 4 lead to SL-T=0, while 5 leads to SL-T=1.

As explained above, IEC 62443 derives fundamental requirements in seven different groups for zones and conduits of a particular IT security architecture. So the result should be a seven-dimensional SL-T vector instead of a scalar value given by (3). But the process description does not give any hint of how to derive the SL-T vector of a zone or conduit from the risk assessment of a threat-vulnerability combination. No explanation is given about how the concept is broken down to the foundational requirements. It may formally be argued that the authors assume that all components of the SL-T vector equal the scalar value derived by (3), but this would, in most cases, lead to very demanding requirements, e.g. for most ICS applications confidentiality is less important than integrity or availability and so the DC foundational requirement can be much weaker than that for SI or RA.

Also, at least for safety-related systems, SL-T=0 does not really make sense as protection against casual or coincidental violation should be provided in any case. It is hard to imagine a system which should not be protected against such threats. For safety-related systems, it is necessary to prevent human errors or foreseeable misuse in any case.

Additionally, there is a difference in the definition of SL between the proposal in IEC 62443-3-2 and the other parts of the standards. By applying formulae (2) and (3), the SL-T is equivalent to a risk reduction, while in the other parts, e.g. 62443-3-3, the SL-T is defined with respect to the attacker type against whom they are to offer protection. The relationship between risk reduction and the type of attacker is not explained, so it is questionable whether the approach fits to other parts of the standard.

The input scales for parameters L and I are ordinal, so we know only the ordering of values 1<2<3<4<5, but have no knowledge about their further relations. For example, we do not know if an impact of 3 is five times more severe than that of 2. We could also re-label the categories to A; B, C, D, E [17].

To make this more tangible, in programming languages such as Pascal or C, such ordinal types could be declared as

```
type
   impact = (trivial, minor, moderate, major, critical);
   likelihood = (remote, unlikely, possible, likely, cer-
   tain);
```

Semantically, only certain operations such as predecessor, successor, ordinal number, greater than, etc., are defined for ordinal data types, but certainly not multiplication or division, which are simply undefined for ordinal data.

What is suggested by Table 1 is that the ordinal data such as "minor" is equated numerically with their order values in their type definition, e.g. `Ord(minor)` which equals 2. These order values are then treated as rational numbers and used without further explanation.

To make this argument clearer, assume that we would have labeled Table 1 with letters instead of numbers. What would $B \cdot C$ mean? Or how would the cyber-security risk reduction factor $B \cdot C/4$ be interpreted? And why should the values be multiplied and not be added?

So we can conclude that the proposal failed due to several reasons, the most important being its insufficient integration with the other parts of the standards and theoretical deficiencies. It also seems a fallacy to quantify risk reduction factors for IT security like practiced in safety.

## 6.2 German DKE 0831-104 proposal.

In a recently published standard for railway automation [18] the approach seems to avoid the uncertainty or infeasibility of credible likelihood estimation, but rather to focus on the capability of the attacker as stipulated by the SL definition. The rationale behind this ap-proach is that the worst case would be a remote attack, which cannot be traced and which has safety impact. This case would deserve the highest IT security requirements.

In a first step, it must be decided whether the zone or conduit is exposed to malicious attacks at all. If no malicious attacks have to be assumed, then SL 1 is assigned for all FR. Otherwise, the parameters already addressed by IEC 62443 would be assessed separately for each zone and each conduit according to Table 1. This means a score is assigned to each of the parameters resources, know-how and motivation of the attacker.

| Score | 2 | 3 | 4 |
|---|---|---|---|
| Resources (RES) | Low | Medium | Extended |
| Know-how (KNO) | Common | System-specific | Extended |
| Motivation (MOT) | Low | Limited | High |

Table 1: Assessment of attacker capability

The following railway specific risk parameters should be considered in addition to the parameters already dealt with in IEC 62443, also in comparison with NIST 800-30:

- Attack location (from where can the attack be launched?)
- Traceability of the attack (in the sense of non-repudiation)
- Potential extent of damage (Safety-critical impact)

It is important that a realistic type of attacker is evaluated, rather than an evaluation of which resources, capabilities and motivation an attacker would need for a successful attack. As it has also been demonstrated that in particular the motivation of an attacker and the location of the attack and its traceability are dependent on one another, the evaluation of the attacker's motivation does not need to take place directly, but is covered by the other parameters.

A combination rule is needed to be able to evaluate these two parameters independently. This is specified in Table 2.

| | R2 | R3 | R4 |
|---|---|---|---|
| K2 | PSL 2 | PSL 3 | PSL 4 |
| K3 | PSL 3 | PSL 3 | PSL 4 |
| K4 | PSL 3 | PSL 4 | PSL 4 |

Table 2 – Preliminary SL assignment

In particular, the following facts were taken into account:

- According to IEC 62443, an SL x is sufficient to successfully ward off an attacker belonging to the combination Rx, Kx and Mx.

- An attacker who possesses higher resources than skills could acquire the applicable skills by using his resources.

- The motivation of an attacker and the location of the attack and its traceability are dependent on one another.

The provisional SL (PSL) listed in Table 6 corresponds to the SL in compliance with EC 62443 without considering railway specific risk factors.

Generally, several attacker types have to be considered. In this case, the highest PSL of the different attacker types' shall be taken into account. The railway specific parameters have a special importance compared with other application domains and can therefore give rise to adapted SL requirements. The PSL may then be altered based on railway-specific risk parameters, e.g.

- location of the asset and the attacker (ORT), e.g. does the attacker need access to the site or can the attack be launched remotely, e.g. from home?
- traceability and non-repudiation of the attack (NAC), e.g. is it possible to trace the attacker and to collect sufficient evidence to identify him?
- potential of the attack (POT), e.g. is there no or limited safety implication of the attack?

All these additional variables are binary and are set to 1 if the question can be answered by YES. The final SL is assignment is then by

$$SL = PSL - \text{maximum}\{ORT, NAC, POT\} \qquad (4)$$

meaning that the PSL can be reduced if there is at least one railway risk reduction factor present, but not more.

It should be noted that, according to IEC 62443, the assessment would have to be carried out for all seven FR. However, from a railway safety point of view, some of the FR have only little safety impact so that FR such as "Data confidentiality" or "Resource availability" should always be assigned SL1 as a default. Also, it can be argued that there is no real reason to distinguish between the other FR, because they are not independent, and it is proposed to allocate the same SL to all five remaining FR. This would lead to only four classes for railway signaling applications

- SL1 =(1,1,1,1,1,1,1)
- SL2 =(2,2,2,1,2,2,1)
- SL3 =(3,3,3,1,3,3,1)
- SL4 =(4,4,4,1,4,4,1)

In the approach presented here, it has to be decided against which kind of attackers the system has to be made secure, which is a decision to be taken by the railway operator and the safety authority. This decision is guided by the parameters in Table 1. The SL then represents the effort which must be made so that the system effectively withstands attacks by these kinds of attackers. Only attackers who exceed this effort considerably might be able to overcome the IT security countermeasures. Different kinds of attackers on railway assets have already been researched [19] and the results were compatible with the classification proposed here.

However it may be criticized that this approach concentrates too much on the attacker capability without exploiting all attack scenarios or taking all security aspects into account. In order to satisfy these critics we extend the approach in the next chapter.

## 7    Combined approach

We can summarize the analysis so far that the approach proposed by IEC 62443-3-2 has several systematic flaws which cannot be easily overcome. In particular, the question of calculating IT security-related risks is very complex and should be avoided [20]. But we have sketched an approach how to derive the SIL from a safety point of view.

However, the use of risk matrices in IT security is so widely used in TRA that it should be kept, but it should be properly used with the definition of SL in IEC 62443.

We start from the following assumptions (without further justification):

- There exists an agreed risk matrix, like figure 2 or 3
- We can derive SLs which are defined by the type of attacker and the measures defined by IEC 62443 (like in the previous chapter)

For the sake of the example, we assume the same sample risk matrix as shown from figure 3 (but we do not use the criticalities). The precise form of the matrix is not important, however there should be a clear procedure which would be followed based on the color code of the results.

In a TRA, we would assess all possible threat scenarios and classify them according to their risk. The following example shows the result for three scenarios X, Y and Z.

Assume we have defined the SL by the type of attacker, say initially SL is equal to (3,3,3,1,3,3,1). Then, we would start the TRA as a check and should arrive at tolerable risks for safety-related threats (usually green or yellow fields, e. g. scenario X or Y). If we arrive at orange or even red classifications (like scenario Z), this means we either misjudged the SL in the beginning or we may have scenarios that represent additional IT security related risks which are not safety-related. For example this might be a loss of reputation after a data breach. This would mean that we have to define additional security requirements which are not safety-related.

# 8 Conclusion

This paper has reviewed the risk assessment approaches with respect to IT security and has proposed a framework for risk assessment with particular focus on railway automation applications. The concept aims at the separation of safety and security aspects, as far as possible. This is achieved by integrating safety-related security requirements into the safety process and the safety case.

It has been explained that this approach matches well with the planned ISA99/IEC62443 series of standards, but that some aspects such as SL allocation need to be adapted. If this adaptation were successful, then railway automation could re-use the ISA99/IEC62443 series, in particular their standardized IT security requirements, and would not have to create its own IT security standards. However, the work presented is still ongoing.

# 9 References

1. http://www.nextgov.com/nextgov/ng_20120123_3491.php?oref=topstory
2. Johnson, C.: CyberSafety: CyberSecurity and Safety-Critical Software Engineering, in: Dale, C. and Anderson, T, (eds.) Achieving System Safety, Proc. 20th Safety-Critical Systems Symposium, Springer, 2012
3. Thomas, M.: Accidental Systems, Hidden Assumptions and Safety Assurance, in: Dale, C. and Anderson, T, (eds.) Achieving System Safety, Proc. 20th Safety-Critical Systems Symposium, Springer, 2012
4. ISA 99, Standards of the Industrial Automation and Control System Security Committee of the International Society for Automation (ISA) on information security, see http://isa99.isa.org/Documents/Forms/AllItems.aspx
5. IEC 62280 Railway applications, Communication, signaling and processing systems –Safety related communication in transmission systems, September 2010
6. IEC 62425 Railway applications, Communication, signaling and processing systems – Safety-related electronic systems for signaling, February 2003
7. ISO/IEC 15408 Information technology – Security techniques – Evaluation criteria for IT security, 2009
8. Common Criteria for Information Technology Security Evaluation, Version 3.1, Revision 3, July 2009. Part 1: Introduction and general model
9. Common Criteria for Information Technology Security Evaluation, Version 3.1, Revision 3, July 2009. Part 2: Functional security components
10. Common Criteria for Information Technology Security Evaluation, Version 3.1, Revision 3, July 2009. Part 3: Assurance security components
11. Commission Implementing Regulation (EU No 402/2013 of 30 April 2013 on the common safety method for risk evaluation and assessment and repealing Regulation (EC) No 352/2009
12. ISO: Information technology - Security techniques - Information security risk management, ISO 27005:2011
13. NIST: Guide for conducting risk assessments, SP800-30, 2012
14. Aven, T. 2011. *Misconceptions of risk*, Wiley, 2010
15. IEC: Industrial communication networks – Network and system security – Part 3-3: System security requirements and security levels, IEC 62443-3-3, 2015
16. IEC 62443-3-2: Security for industrial automation and control systems – Part 3-2: Security risk assessment and system design, draft for comments, August 2015

17. Wikipedia: Ordinal Scale, https://en.wikipedia.org/wiki/Level_of_measurement#Ordinal_scale, last accessed on 2016-10-25
18. Electric signaling systems for railways – Part 104: IT Security Guideline based on IEC 62443 (in German), 2015
19. Schlehuber, C.: Analysis of security requirements in critical infrastructure and control systems (in German), Master thesis, TU Darmstadt, 2013
20. Braband, J., Schäbe, H.: Probability and Security – Pitfalls and Chances: in Proc. Advances in Risk and Reliability Technology Symposium 2015, Loughborough, 2015